\documentclass[prd,twocolumn,floatfix,altaffilletter,superscriptaddress,floatfix,
               tightenlines,showpacs,showkeys,preprintnumbers,nofootinbib]{revtex4-1}

\usepackage[colorlinks=true,citecolor=blue,linkcolor=blue]{hyperref}
\usepackage{amsmath,amssymb}
\usepackage{epsfig}  
\usepackage{graphicx}               
\usepackage{url}
\usepackage{hyperref}
\usepackage{color}
\usepackage{multirow}
\usepackage{placeins}
\usepackage[dvipsnames]{xcolor}
\usepackage{dcolumn}

\usepackage{tikz}
\usetikzlibrary{trees}
\usetikzlibrary{decorations.pathmorphing}
\usetikzlibrary{decorations.markings}

\definecolor{jblue}  {RGB}{20,50,100}
\definecolor{npurple}  {RGB} {153, 51, 204}  
\definecolor{wred}   {RGB}{217,0,56}
\definecolor{white}   {RGB}{255,255,255}

\definecolor{korange}   {RGB}{235, 80,  43}
\definecolor{korange2}   {RGB}{245, 100,  63}
\definecolor{kyelloworange}   {RGB}{255, 210,  110}
\definecolor{kyelloworange2}   {RGB}{240, 170,  90}
\definecolor{kred}   {RGB}{204,  102, 153}
\definecolor{kpurple}   {RGB}{153,  61, 190}
\definecolor{kpurplelight}   {RGB}{213,  161, 230}

\usepackage{cleveref}
\usepackage{subfigure} 
\usepackage{lipsum}
\definecolor{red}{rgb}{1.0, 0, 0}
 \usepackage{gensymb}
\allowdisplaybreaks

\newcommand{\ev}[1]{\ensuremath{\left\langle #1 %
                     \right\rangle}} 

\tikzset{
  gaugeboson/.style={decorate,deblueation={snake, amplitude = 6pt, post length = 1 pt, pre length = 4 pt},draw=magenta},
  fermion/.style={draw=black,postaction={decorate},decoration={markings,mark=at position .55}},
  fermionin/.style={draw=black,postaction={decorate},decoration={markings,mark=at position .55 with {\arrow[draw=black]{<}}}},
  fermionout/.style={draw=black,postaction={decorate},decoration={markings,mark=at position .55 with {\arrow[draw=black]{>}}}},
  gluon/.style={decorate,draw=magenta,decoration={coil,amplitude = 6pt,segment length=8pt}},
  connect/.style={draw=black,postaction={decorate},decoration={markings}}, 
  gluon2/.style={decorate,draw=magenta,decoration={coil,amplitude = 3pt,segment length=4pt}},
  gaugeboson2/.style={decorate,decoration={snake, amplitude = 3pt, segment length=6pt},draw=black}
}
        
\tikzset{
  photon/.style={decorate, decoration={snake}, draw=npurple,very thick},
  boson/.style={decorate, decoration={snake}, draw=npurple,very thick},
  electron/.style={draw=jblue,very thick, postaction={decorate},
           decoration={markings,mark=at position .55 with {\arrow[draw=jblue]{>}}}
  },
  electron2/.style={draw=jblue,very thick, postaction={decorate},
           decoration={markings,mark=at position .55 with {\arrow[draw=jblue]{<}}}
  },
  fermion/.style={draw=jblue,very thick, postaction={decorate},
            decoration={markings,mark=at position .55 with {\arrow[draw=jblue]{}}}
  },
  gluon/.style={decorate, draw=korange,very thick, 
    decoration={coil,amplitude=4pt, segment length=6pt}},
  higgs/.style={draw=wred,very thick, postaction={decorate},
           decoration={markings,mark=at position .55 with {\arrow[draw=wred]{>}}}
  },
  nothing/.style={draw=white,very thick}
}


\begin{document}

\title{Production of keV Sterile Neutrinos in Supernovae:
       New Constraints and Gamma Ray Observables}

\author{Carlos A.\ Arg\"uelles} \email{caad@mit.edu}
\affiliation{Massachusetts Institute of Technology, Cambridge, MA 02139, USA}
\affiliation{Wisconsin IceCube Particle Astrophysics Center, Madison, WI 53703, USA}
\author{Vedran Brdar} \email{vbrdar@uni-mainz.de}
\author{Joachim Kopp} \email{jkopp@uni-mainz.de}
\affiliation{PRISMA Cluster of Excellence, 55099 Mainz, Germany and \\
             Mainz Institute for Theoretical Physics,
             Johannes Gutenberg-Universit\"{a}t Mainz, 55099 Mainz, Germany}

\preprint{MITP/16-034}

\begin{abstract}
We study the production of sterile neutrinos in supernovae, focusing in particular
on the keV--MeV mass range, which is the most interesting range if sterile neutrinos
are to account for the dark matter in the Universe. Focusing on the simplest scenario
in which sterile neutrinos mixes only with muon or tau neutrino, we
argue that the production of keV--MeV sterile neutrinos can be strongly enhanced by
a Mikheyev--Smirnov--Wolfenstein (MSW) resonance, so that a substantial flux
is expected to emerge from a supernova, even if vacuum mixing angles between
active and sterile neutrinos are tiny.  Using energetics arguments, this yields
limits on the sterile neutrino parameter space that reach down to mixing
angles on the order of $\sin^2 2\theta \lesssim 10^{-14}$ and are up to an order of
magnitude stronger than those from X-ray observations.
While supernova limits suffer from larger systematic uncertainties than X-ray limits
they apply also
to scenarios in which sterile neutrinos are not abundantly produced in the early
Universe.
We also compute the flux
of $\mathcal{O}(\text{MeV})$ photons expected from the decay of sterile neutrinos
produced in supernovae, but find that it is beyond current observational reach even
for a nearby supernova.
\end{abstract}

\maketitle

\section{Introduction}
\label{sec:intro}

One of the most auspicious candidate particles for the dark matter in the
Universe is the sterile neutrino---an electrically neutral fermion with a mass on
the order of keV--MeV that couples to ordinary matter only through a tiny mass mixing
with Standard Model (SM) neutrinos~\cite{Abazajian:2001nj, Kusenko:2009up}.
In the simplest sterile neutrino scenarios, it is assumed that the abundance of
sterile neutrinos $\nu_s$ is zero at the end of inflation, and they are later
produced through their mixing with SM (active) neutrinos
$\nu_a$~\cite{Dodelson:1993je, Shi:1998km} (see also \cite{Petraki:2007gq,
Merle:2013wta, Merle:2015vzu}).
Experimental constraints on the mass of keV sterile neutrinos and their
mixing with SM neutrinos arise from the measured DM relic
density~\cite{Ade:2015xua, Vincent:2014rja}, from Pauli blocking (the
Tremaine--Gunn bound)~\cite{Tremaine:1979we, Iakubovskyi:2013},
from Lyman-$\alpha$ forests~\cite{Baur:2015jsy, Merle:2014xpa, Schneider:2016uqi}, and
from X-ray searches for radiative decays of sterile neutrinos $\nu_s \to \nu_a
+ \gamma$~\cite{Horiuchi:2013noa, Boyarsky:2005us, Abazajian:2006jc, Abazajian:2009hx,
Bulbul:2014sua, Sekiya:2015jsa}. From the combination of these constraints, one concludes that
the $\nu_s$ mass should be $m_s \gtrsim 4$~keV, and its mixing angle with the
SM neutrinos should be $\sin^2 2\theta \lesssim 10^{-6}$ in the simple two-flavor
approximation.

In this letter, we add a new limit to this inventory of constraints by
considering sterile neutrino production in core-collapse supernovae
(SN)~\cite{
  Mikheev:1987hv, 
  Kainulainen:1990bn, 
  Raffelt:1992bs, 
  Shi:1993ee,  
  Nunokawa:1997ct, 
  Hidaka:2006sg, 
  Hidaka:2007se, 
  Raffelt:2011nc, 
  Wu:2013gxa, 
  Warren:2014qza, 
Warren:2016slz}. 
A supernova develops when a $\gtrsim 9 M_\odot$ star runs out of nuclear fuel.
The thermal pressure that normally counteracts gravity disappears, and the core
of the star collapses into a neutron star.  The temperature in the nascent
neutron star is $\gtrsim \text{MeV}$, so that a thermal population of (active)
neutrinos is produced.  These $\nu_a$ can oscillate into $\nu_s$, which
escape the exploding star unhindered and may carry away significant amounts of
energy~\cite{Raffelt:1990yz,Shi:1993ee,Raffelt:2011nc}.  Constraints on
anomalous energy loss from SN~1987A will thus allow us to constrain the sterile
neutrino parameters. Since $\nu_a \to \nu_s$ conversion can be resonant
thanks to the ultrahigh matter density $\sim 10^{14}\;\text{g/cm}^3$ in the SN
core, these limits will be very strong.  The flux of sterile neutrinos with
$\mathcal{O}(\text{MeV})$ energies escaping from a supernova leads to
a flux of secondary gamma rays when they decay, and we study this flux
as well.

\begin{figure}
   \centering
   \begin{tabular}{c}
     \includegraphics[width=\columnwidth]{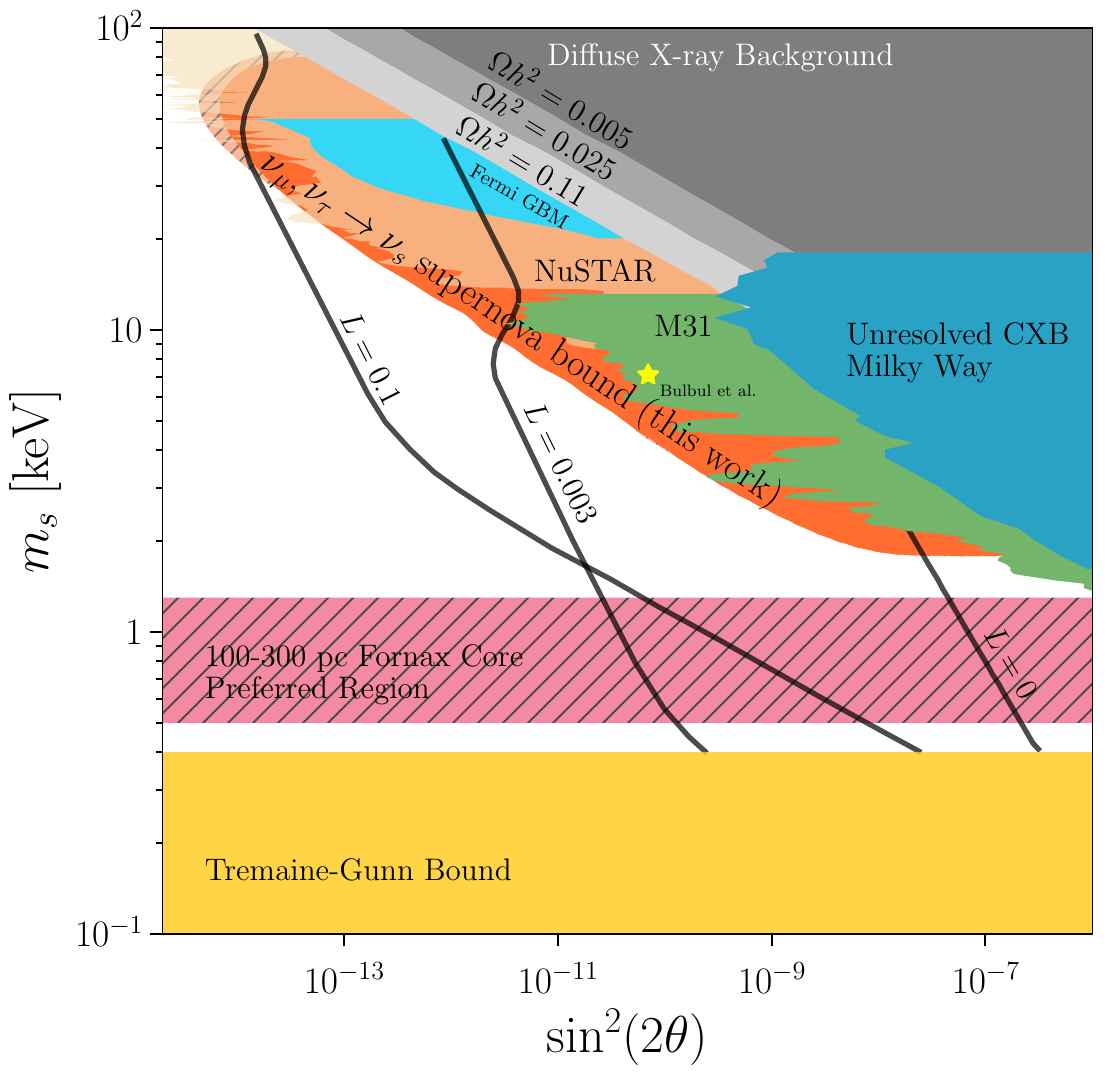}
   \end{tabular}
   \caption{Supernova bounds on the sterile neutrino mass $m_s$ and mixing
     $\sin^2 2\theta$ (orange, this work) between $\nu_x$ $(x=\mu$ or $\tau)$ and $\nu_s$ flavors
     compared to previous constraints~\cite{Abazajian:2006jc, Abazajian:2009hx,
     Bulbul:2014sua} from the Tremaine--Gunn bound
     \cite{Tremaine:1979we,Iakubovskyi:2013} (yellow), from NuSTAR
     observation of Galactic center \cite{Perez:2016tcq} (pale orange),
     from X-ray searches
     in the Andromeda Galaxy M31~\cite{Horiuchi:2013noa} (green), from the Fermi GBM
     all-sky analysis~\cite{Ng:2015gfa} (purple), and from the
     galactic~\cite{Abazajian:2006jc} (dark blue) and
     extragalactic~\cite{Boyarsky:2005us,Abazajian:2009hx,Bulbul:2014sua} (gray)
     diffuse X-ray background. For the latter constraint, we show also how it is
     modified if sterile neutrinos account for only a fraction of the DM in the
     Universe.  It is understood that the other X-ray limits would shift in a
     similar way if $\Omega h^2 < 0.11$.  The hatched pink region indicates the region
     preferred by the properties of the core of the Fornax dwarf
     galaxy~\cite{Strigari:2006ue}, whereas a yellow star shows the parameter point
     that could explain the 3.5~keV X-ray line hinted at by
     refs.~\cite{Bulbul:2014sua, Boyarsky:2014jta}.  Black curves illustrate the
     parameter regions in which the Dodelson--Widrow
     mechanism~\cite{Dodelson:1993je} ($L=0$) or the Shi--Fuller
     mechanism~\cite{Shi:1998km} with lepton assymetry $L > 0$ would yield the
     correct DM relic density $\Omega h^2 = 0.11$. The hatched band at large
     $m_s$ indicates the theoretical uncertainty in our limit stemming from our
     treatment of energy transport inside the supernova (see \cref{sec:energy_loss}).
   }
   \label{fig:limit-comparison}
\end{figure}

Our main results are summarized in
\cref{fig:limit-comparison}.  The solid orange
exclusion region shows that, in the mass range
$m_s \sim 2$--$80$~keV, limits from energy loss in supernovae surpass previous
limits by up to two orders of magnitude in $\sin^2 2\theta$ for the case when 
$\nu_s$ mixes with $\nu_\mu$ or $\nu_\tau$.
Note that, unlike the other limits shown in
\cref{fig:limit-comparison}, our bounds would still
hold if sterile neutrinos are not part of the DM in the Universe.  In the
following, we discuss in detail how we have obtained our new limits and
sensitivity estimates.

\section{Sterile neutrino production in Supernovae}
\label{sec:sterile-production} 

We consider a simplified two-flavor oscillation picture with mixing between a
sterile neutrino\footnote{In this work, for brevity, the term ``sterile neutrino" is denoting both sterile neutrinos and antineutrinos. The only exception is when discussing the adiabatic production where we refer explicitly to ``sterile antineutrinos" because the resonance appears in the antineutrino channel.}
 $\nu_s$ and one species of active neutrinos $\nu_x, \, x=\mu
\,\, \text{or} \,\, \tau$. We do not consider mixing between $\nu_s$ and electron
neutrinos to avoid complications arising from
charged current interactions between $\nu_e$ and electrons/positrons. The
flavor basis Hamiltonian describing neutrino propagation in matter includes the
vacuum oscillation term and a Mikheyev--Smirnov--Wolfenstein (MSW) potential $V_x$
that describes coherent forward scattering of $\nu_x$ on
the background matter via $Z$ exchange~\cite{Wolfenstein:1977ue,
Mikheev:1986gs,Mikheev:1986wj}:
\begin{align}
 H = \frac{\Delta m^2}{4E}
     \begin{pmatrix}
        - \cos\, {2 \theta}  &  \sin\, {2 \theta} \\
          \sin\, {2 \theta}  &  \cos\, {2 \theta}     
     \end{pmatrix} +
     \begin{pmatrix}
        V_{x} &   0 \\
         0    &    0      
     \end{pmatrix} .
  \label{eq:Hamilton}
\end{align}
Here, $\theta \ll 1$ is the $\nu_{s}$--$\nu_{x}$ mixing angle in vacuum and
$\Delta m^2 \simeq m_s^2$ is the mass squared difference between the two mass
eigenstates \footnote{In this paper, we refer to $m_s$ as the sterile
  neutrino mass, even though we mean of course the mass of the heavy mass
eigenstate which is mostly a $\nu_s$, with only a tiny admixture of $\nu_x$.}.
The MSW potential is   $V_x = \pm \sqrt{2} G_F (-N_n/2 + N_{\nu_e} -
N_{\bar{\nu}_e})$, where $G_F$ is the Fermi constant and $N_{n}, \,
N_{\nu_{e}}, \, \text{and} \, N_{\bar{\nu}_{e}}$ are the neutron, electron
neutrino, and electron antineutrino number densities. The $+$ ($-$) sign
corresponds to the potential experienced by neutrinos (antineutrinos).  Since
$|N_{\nu_e} - N_{\bar\nu_e}| \ll N_n/2$, the potential is positive for
antineutrinos and negative for neutrinos.  Note that we neglect terms that
would arise from differences between the $\nu_x$ and $\bar\nu_x$ number
densities because such differences are expected to be small in the parameter
region where our limits will lie~\cite{Raffelt:1990yz}.  We have checked
that even adding an MSW potential corresponding to a maximal asymmetry does
not alter our results significantly.  We also neglect the momentum and angular
dependence of neutrino self-interactions \cite{Duan:2010bg, Dasgupta:2007ws}
and instead restrict ourselves to the simplified formalism used in
\cite{Raffelt:2011nc,Nunokawa:1997ct,Shi:1993ee}.  Since $N_n$, $N_{\nu_e}$,
and $N_{\bar\nu_e}$ are extremely large in the supernova core and gradually
decrease with radius, most antineutrinos will encounter an MSW resonance on
their way out of the exploding star. At the resonance,
\begin{align}
  \cos 2\theta = 2 V_x E_\text{res} / m_s^2 \,,
  \label{eq:msw}
\end{align}
and the effective mixing angle $\theta_m$ in matter becomes
maximal~\cite{Wolfenstein:1977ue,Mikheev:1986gs,Mikheev:1986wj,Akhmedov:1999uz}.

We consider two physically different mechanisms for sterile neutrino production in
supernovae:

\emph{(i) Adiabatic flavor conversion at an MSW resonance.} A $\bar{\nu}_x$ of energy
$E$ streaming away from the supernova core can convert to sterile antineutrino $(\bar{\nu}_s)$
when the matter density, and thus the MSW potential $V_x$, has reached the value
satisfying the resonance condition, \cref{eq:msw}~\cite{Mikheev:1986if}.
Each point $(t,R)$ in time
and space corresponds to a specific value of $V_x$; therefore, antineutrinos of a
specific energy $E_\text{res}(t,R)$ are resonantly converted at this point.
We take the local antineutrino luminosities
and spectra and the local matter densities from the simulation of an $8.8 M_\odot$
supernova by the Garching group~\cite{Huedepohl2010} (see also
\cite{Rampp:2002bq,Liebendoerfer:2003es,Marek:2005if,Buras:2005rp}).

Hard scattering processes must be rare to give antineutrinos sufficient time to
convert adiabatically.  We therefore require the spatial width of the MSW
resonance regions,
\begin{align}
  R_{\text{width}} = 2 \sin 2\theta \bigg( \frac{1}{V_x} \,
                      \frac{\partial V_x}{\partial R} \bigg)^{-1} \,,
  \label{eq:reswidth}
\end{align}
to be smaller than the mean free path $\lambda_\text{mfp}$.
The number $d\mathcal{N}_s^\text{MSW} / dE$ of $\bar{\nu}_s$
in an energy interval $[E, E+dE]$ produced by adiabatic
flavor conversion at the MSW resonance is given
by~\cite{Shi:1993ee}
\begin{align}
  \frac{d\mathcal{N}_s^\text{MSW}}{dE}
    &= \int_0^t \! dt' \, 4\pi R_\text{res}^2 \,
         n_\nu(t',R_\text{res}) \,             
                              \nonumber\\
    &\qquad\times
         f_x(E) \, P_\text{res}(E) \,
         \frac{E^2}{\bar{E}^3} \,
         \Theta(\lambda_\text{mfp} - R_{\text{width}}) \,.
  \label{eq:dNdE-MSW}
\end{align}
Here, the time integral runs from the time of core bounce ($t'=0$) until $\sim
9$~sec later, and $R_\text{res}(t', E)$ is the radius at which the resonance
energy is $E$ at time $t'$.  The quantity $n_\nu(t',R_\text{res})$ is the
active (anti)neutrino number density at time $t'$ and radius $R_\text{res}$, $f_x(E)$
is the energy distribution of active (anti)neutrinos, $\bar{E}$ is the average
(anti)neutrino energy, $P_\text{res}(E)$ is the flavor conversion probability at the
resonance, and the Heaviside $\Theta$ function implements the condition that
antineutrinos must have enough time between collisions to convert adiabatically.
It is crucial that antineutrinos do not encounter more than one MSW resonance on
their way out of the supernova.  (This would be different if we considered
mixing between $\nu_s$ and $\nu_e$ instead of
$\nu_{\mu,\tau}$~\cite{Hidaka:2006sg, Hidaka:2007se}.) Note that we can make
the strongly simplifying assumption of radial symmetry, and we also neglect the
depletion of active antineutrinos by conversion into $\bar{\nu}_s$. Moreover, we
do not need to consider $\bar{\nu}_x$ streaming inwards. They would convert to
$\bar{\nu}_s$ at the resonance, then travel through the core, and convert back to
$\bar{\nu}_x$ on its far side.

We parameterize $f_x(E)$ as~\cite{Keil:2002in}
\begin{align}
  f_x(E) &= \frac{(1+\alpha)^{3+\alpha}}{\Gamma(3 + \alpha)}
           \biggl( \frac{E}{\bar{E}} \biggr)^\alpha
           \exp\biggl[ -(1 + \alpha) \frac{E}{\bar{E}} \biggr] \,,
  \label{eq:fx}
\end{align}
with normalization
$\int_0^\infty \! dE \, E^2 \, f_x(E) = \bar{E}^3$.
(This relation defines $\bar{E}$.) The ``pinching parameter''
$\alpha$ desribes the degree to which $f_x(E)$
differs from a Maxwell--Boltzmann distribution.

The $\bar{\nu}_x \to \bar{\nu}_s$ conversion probability at the resonance
is given by the Landau--Zener formula~\cite{Nunokawa:1997ct,
Shi:1993ee, Akhmedov:1999uz}
\begin{align}
  P_\text{res}(E) = 1 - \exp\biggl[-\frac{\pi^2}{2}
    \frac{R_\text{width}}{L_\text{osc}^\text{res}} \biggr] \,, 
  \label{eq:P-res}
\end{align}
with the oscillation length at the resonance, $L_\text{osc}^\text{res}
\simeq 2 \pi / (V_x \, \sin 2\theta)$.

\begin{figure}
   \centering
   \includegraphics[width=0.4\textwidth]{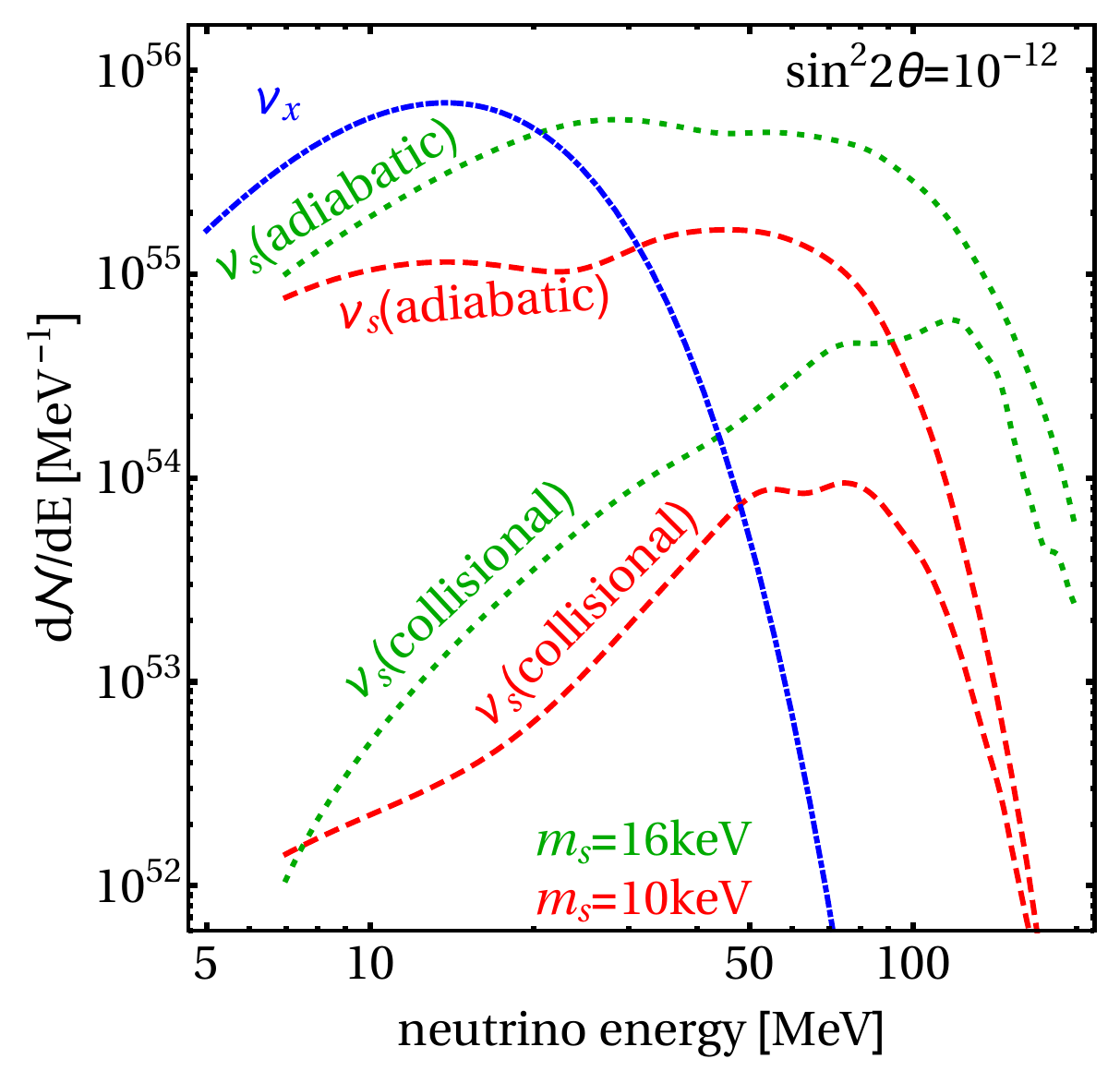}
   \caption{Time integrated spectra of active neutrinos (blue) and sterile neutrinos (green and red) produced in an $8.8 M_\odot$ supernova~\cite{Huedepohl2010} during the first $\sim 9$ seconds after the core bounce. 
     Contributions from
     adiabatic flavor conversion and collisional production
     are shown separately. The $\nu_s$ spectrum extends to much
     higher energies than the $\nu_x$ spectrum because
     flavor conversion occurs in a region where temperatures are much higher
     than at the last scattering surface of $\nu_x$.
   }
   \label{fig:spectrum}
\end{figure}

We find that adiabatic flavor conversion occurs mostly at radii $\sim
10$--$15$~km, still inside the neutrino sphere at $\sim 20$--$30$~km.
In \cref{fig:spectrum}, we compare $d\mathcal{N}_s^\text{MSW} / dE$
to the spectrum of active neutrinos.
We see that the sterile antineutrino spectrum extends to
higher energies because most of the flavor conversion happens in a
high-temperature region from which $\nu_s$ can stream out
freely, while $\nu_x$ are still trapped.

\emph{(ii) Collisional production.}
Sterile neutrinos can also be produced from a mixed $\nu_x$--$\nu_s$ state in
hard scatterings on nucleons, electrons, and positrons.  We take the interaction
rate $\Gamma_x$ to be approximately equal to the dominant scattering rate on
neutrons, with the corresponding cross section given by~\cite{Giunti:2007ry,ultimate-nu}
$\sigma_{\nu n} \simeq G_F^2 E_\nu^2 (1 + g_A^2 ) / 4 \pi$, with $g_A \simeq 1.23$.
The physical picture for collisional $\nu_s$ production is as follows:
starting with an ensemble of only $\nu_x$ at $t=0$, each of them soon
acquires a small $\nu_s$ admixture $\propto
\sin^22\theta_\text{m}$ by oscillation.  A collision causes the
collapse of the resulting mixed state into either a pure $\nu_x$ or a pure
$\nu_s$.
Afterwards, oscillations start anew, and $\nu_x$ are quickly
replenished. After many collisions, the $\nu_s$ abundance
is proportional to $\sin^22\theta_{\text{m}}$ and $\Gamma_x$.

Up to a factor of $1/2$, this intuitive picture leads to the correct quantum
mechanical Boltzmann equation \cite{Stodolsky:1986dx,Raffelt:1992bs,
Dodelson:1993je,Abazajian:2001nj,Thomson}
\begin{align}
  \frac{\partial}{\partial t} \frac{dn_s}{dE}
  = \frac{\Gamma_x}{2} \, \ev{P_{\nu_x \rightarrow \nu_s}} \, \frac{dn_x}{dE} \,.
  \label{eq:boltzmann}
\end{align}
Here, $dn_s/dE$ and $dn_x/dE$ are the energy spectra of sterile
and active neutrinos, respectively. 
At any given spacetime point, $dn_x(t,R,E)/dE$
is related to the distribution function in \cref{eq:fx} via
$(1/n_x) dn_x/dE = (E^2 / \bar{E}^3) \, f_x(E)$.
The averaged oscillation probability $\ev{P_{\nu_x \rightarrow \nu_s}}$
is~\cite{Abazajian:2001nj}
\begin{align}
  \ev{P_{\nu_x \rightarrow \nu_s}}
    &= \frac{1}{2}
         \frac{\sin^2 2 \theta}{(\cos 2\theta - 2 V_x E / m_s^2)^2
             + \sin^2 2\theta + (\Gamma_x E / m_s^2)^2} \,.
  \label{eq:boltzmann2}
\end{align}
The extra term $(\Gamma_x E / m_s^2)^2$ in the denominator compared to the
usual expression for the mixing angle in matter~\cite{Akhmedov:1999uz} accounts
for the suppression of $\nu_s$ production when $\lambda_\text{mfp}$ is much
smaller than the oscillation length, so that oscillations do not have
time to develop between collisions (quantum Zeno effect).

Integrating \cref{eq:boltzmann} over time and radius leads to the
energy spectrum of sterile neutrinos produced collisionally,
\begin{align}
  \frac{d\mathcal{N}_s^\text{coll}}{dE}
    &= \int_0^t \! dt' \int_0^{R_\text{core}} \!\!\! dR' \, 4 \pi R'^2
         \frac{\Gamma_x}{2} \ev{P_{\nu_x \rightarrow \nu_s}} \frac{dn_x}{dE} \,.
  \label{eq:boltzmann6}
\end{align}
We again evaluate $d\mathcal{N}_s / dE$ numerically
using the data from \cite{Huedepohl2010}.  The resulting $\nu_s$ spectra, shown in
\cref{fig:spectrum}, can be harder than the ones from adiabatic
production because the collisional production rate depends
on $\Gamma_x$, which grows proportional to $E^2$.

\section{Constraints from Supernova Luminosity}
\label{sec:luminosity-constrain} 

We can constrain the energy output in sterile neutrinos from SN~1987A by
comparing the observed energy output in active neutrinos of $\mathcal{E}_a =
\text{few} \times 10^{53}$~ergs~\cite{Loredo:2001rx,Pagliaroli:2008ur} to the
gravitational energy released in the collapse of a stellar core at the
Chandrasekhar mass, which is also on the order of $\mathcal{E}_\text{tot} =
\text{few} \times 10^{53}$~ergs~\cite{Huedepohl2010,Zuber:2011}.  If a
substantial fraction of $\mathcal{E}_\text{tot}$ was carried away by sterile
neutrinos, the observed $\mathcal{E}_a$ could not be explained \footnote{A
  similar constraint could also be obtained by considering the observed
  duration of the neutrino emission from SN~1987A.  Efficient production of
  sterile neutrinos would expedite the cooling of the supernova core,
  shortening the neutrino burst~\cite{Raffelt:1990yz, Raffelt:2011nc}.}.
We therefore consider the ratio $\mathcal{R}(\sin^2 2\theta,m_s) \equiv
\mathcal{E}_s(\sin^2 2\theta,m_s) / \mathcal{E}_\text{tot}$~\cite{Mikheev:1987hv}.
We assume that
$\mathcal{R}$ depends only weakly on the mass and type of the progenitor star,
so that the values obtained for the supernova simulated in \cite{Huedepohl2010}
are a good proxy for SN~1987A. 
Indeed, the energy emitted in neutrinos from the simulated supernova 
is roughly $1.5\cdot 10^{53}$~erg \cite{Huedepohl2010},
which is of the same order of magnitude as in Supernova 1987A.

Our computation of $\nu_s$ production is only
self-consistent for $\mathcal{R} \ll 1$ because we neglect depletion of active neutrinos.
Extrapolating it to larger values nevertheless and setting a limit by
requiring $\mathcal{R} < 1$ yields the leftmost edge of the hatched
uncertainty band in \cref{fig:limit-comparison}. To obtain the right edge of
the band, we have also implemented a more robust treatment of
active neutrino fluxes that parameterizes both depletion due to $\nu_s$
production and replenishment from diffusion (see \cref{sec:energy_loss} for
details).

\begin{figure}
  \centering
  \begin{tabular}{cc}
    \includegraphics[width=0.5\columnwidth]{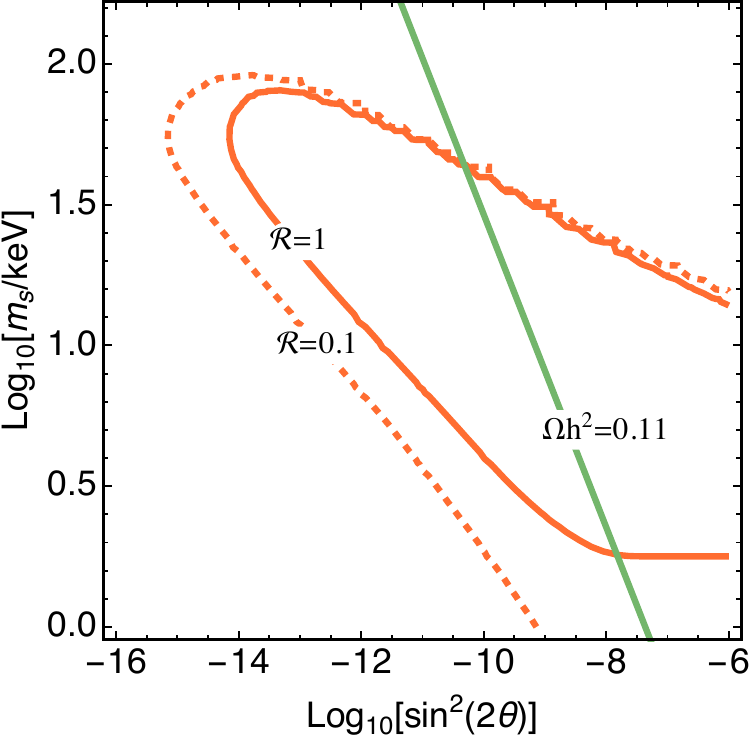} &
    \includegraphics[width=0.5\columnwidth]{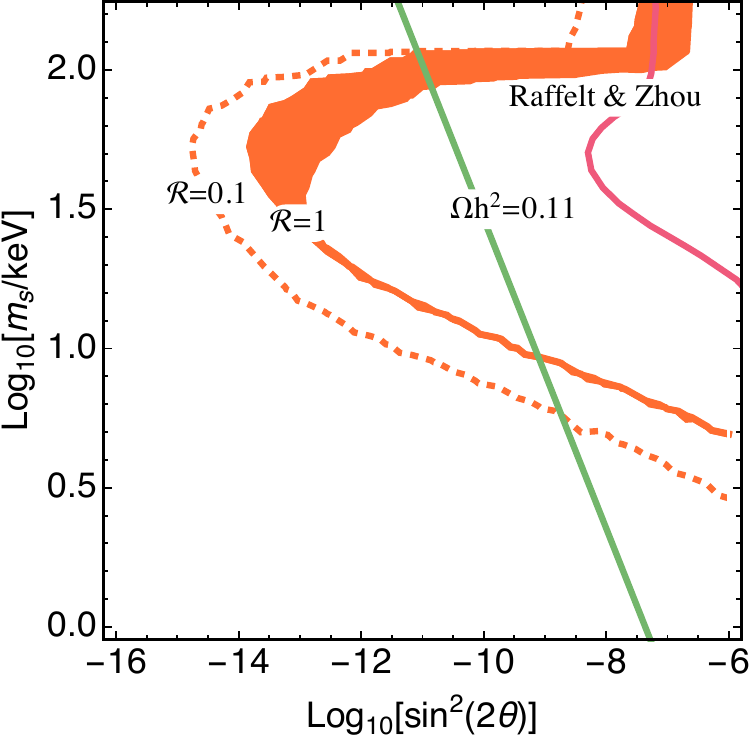} \\
    (a) & (b)
  \end{tabular}
  \caption{Constraints on the sterile neutrino parameter space from energy loss
    in supernovae, considering (a) adiabatic $\nu_x \to \nu_s$ conversion only
    or (b) collisional $\nu_s$ production only. The
    solid and dotted orange curves correspond to different assumptions on the
    maximum allowed energy loss, expressed here in terms of the
    ratio $\mathcal{R}$ of the energy output in sterile neutrinos and the total
    energy output. For comparison, we also show in green
    the parameters for which the Dodelson--Widrow mechanism~\cite{Dodelson:1993je}
    predicts the correct relic abundance of $\nu_s$. In panel (b) for $\mathcal{R} = 1$,
    the shaded
    region illustrates the theoretical uncertainty in our limit stemming from our
    treatment of energy transport inside the supernova (see \cref{sec:energy_loss}).
  }
  \label{fig:energy-constraint}
\end{figure}

Our limits on the $\nu_s$ parameter space are shown in
\cref{fig:energy-constraint} for adiabatically and collisionally produced
$\nu_s$ separately, and in \cref{fig:limit-comparison} for the combination of
both production mechanisms (requiring $\mathcal{R} < 1$). Thanks to MSW
enhancement, our constraints reach down to $\sin^2 2\theta \sim 10^{-14}$ at
$m_s \sim 10$--$100$~keV, surpassing all other limits in this mass range.  The
strength of our bounds can be understood by noting that according to
\cref{eq:boltzmann2}, strong resonant conversion is possible down to
$\sin^2 2\theta \sim 10^{-14}$. At smaller values, the damping term $(\Gamma_x
E/m_s^2)^2$ in the denominator suppresses $\ev{P_{\nu_x \to \nu_s}}$ at
resonance. Moreover, considering the smooth density profiles from
\cite{Huedepohl2010} rather than assuming a constant density core as e.g.\ in
ref.~\cite{Raffelt:2011nc} implies that, in the mass range between 10--100~keV,
neutrinos of any energy will experience a resonance somewhere along their
trajectory.
It is primarily due to this effect that the limits on $\sin^2 2\theta$
presented here are (in the most sensitive region around $m_s =50$ keV) about
$3$--$4$ orders of magnitude stronger than those in ref.~\cite{Raffelt:2011nc}.

The shape of the exclusion regions in \cref{fig:energy-constraint} can be
understood as follows.  For adiabatic conversion at small $m_s$, the
oscillation length at the MSW resonance $L_\text{osc}^\text{res}$ is large,
making flavor conversion non-adiabatic according to \cref{eq:P-res}.  At large
$m_s \gtrsim 100$~keV, the resonance condition of \cref{eq:msw} cannot be
satistifed. At $m_s \sim
\text{few} \times 10$~keV, adiabatic flavor conversion is most effective at low
$\sin^2 2\theta \sim 10^{-14}$.  For larger $\sin^2 2\theta$, the radial width
$R_\text{width}$ of the resonance region becomes too large, so that neutrinos
scatter before having a chance to convert.  Also collisional production is most
effective when the mixing angle in matter is MSW-enhanced. At $m_s \lesssim
10$~keV, the resonance condition of \cref{eq:msw} is fulfilled only in the
outer regions of the supernova, where the scattering rate is too low for
effective collisional production.  At $m_s \gtrsim 100$~keV, the resonance
condition is never satisfied.

\section{Photon flux from $\nu_s \to \nu_a \gamma$}
\label{sec:photon-constrain} 

Sterile neutrinos, once produced, decay to photons via $\nu_s \to \nu_a
\gamma$, a process mediated by loop diagrams involving charged leptons and
$W$ bosons.  The decay rate is~\cite{Pal:1981rm,Xing:2011}
\begin{align}
  \Gamma_{\nu_s \to \nu_a \gamma} = 1.38 \cdot 10^{-29}\;\text{sec}^{-1}
    \biggl(\! \frac{\sin^2 2\theta}{10^{-7}} \biggr)
    \biggl(\! \frac{m_s}{\text{1 keV}} \biggr)^5\!.
  \label{eq:photon}
\end{align}
Thus, if $\nu_s$ are abundantly produced
in a supernova, we expect the explosion to be accompanied by a flux of energetic
($\mathcal{O}(\text{1--100~MeV})$) secondary gamma rays.  Photons in this energy
range are not normally expected from a supernova or supernova remnant,
both of which emit X-rays only at
energies $\lesssim 10$~keV~\cite{Soderberg:2008uh,Vink:2012}.
The arrival times of the gamma rays
from $\nu_s$ decay are spread out over a time interval
\begin{align}
  \Delta t
    &\simeq 3.6~\text{hrs} \times
              \bigg( \frac{d}{1~\text{kpc}} \bigg)
              \bigg( \frac{m_s}{1~\text{keV}} \bigg)^2
              \bigg( \frac{\text{1~MeV}}{E_\gamma} \bigg)^2 \,,
  \label{eq:Delta-t}
\end{align}
where $d$ is the distance to the supernova and $E_\gamma$ is the gamma ray energy.
Sterile neutrinos decaying immediately after their production lead to gamma
rays that reach the Earth at the same time as the active neutrino burst. Gamma rays
from $\nu_s$ decaying only after travelling a distance $\simeq d$ are delayed
by $\Delta t$.

\begin{figure}
   \centering
   \includegraphics[width=0.7\columnwidth]{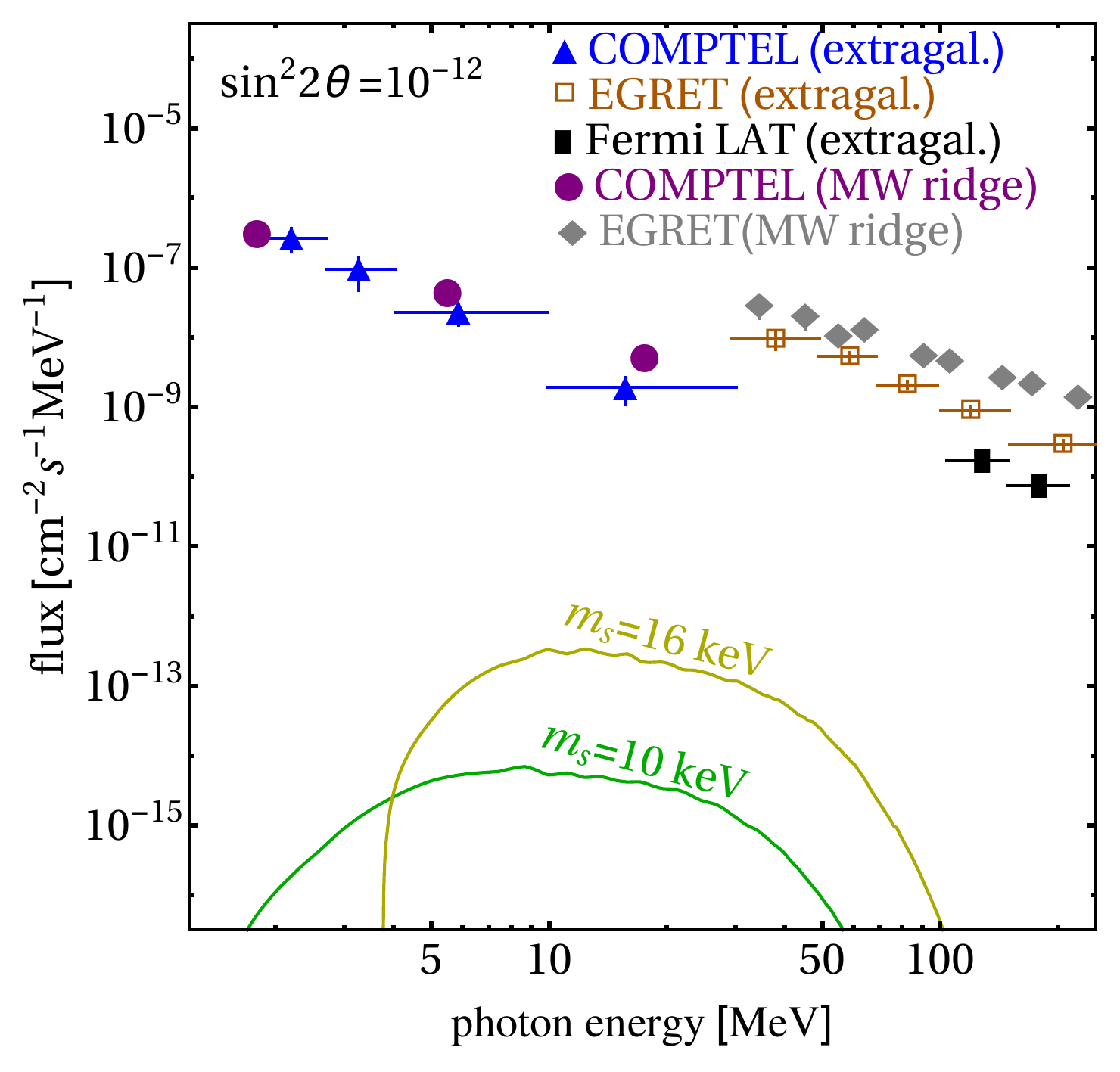}
   \caption{Photon flux from the decay of sterile neutrinos produced in a
     nearby ($d = 1$~kpc) supernova for two different benchmark points in the
     $m_s$--$\sin^2 2\theta$ plane. We compare to observed spectra of diffuse gamma rays
     from COMPTEL~\cite{Strong:1994,Kappadath:1996,Kinzer:1999}, EGRET~\cite{Hunter:1997,
     Sreekumar:1997un, Sreekumar:1997yg}, and Fermi-LAT~\cite{Ackermann:2014usa},
     assuming angular bin sizes of 1~degree for COMPTEL, 5~degrees for EGRET, and 3~degrees
     for Fermi-LAT~\cite{Schoenfelder2013}.
   }
   \label{fig:photon-spectrum}
\end{figure}

\begin{figure}
   \centering
   \begin{tabular}{c}
     \includegraphics[width=\columnwidth]{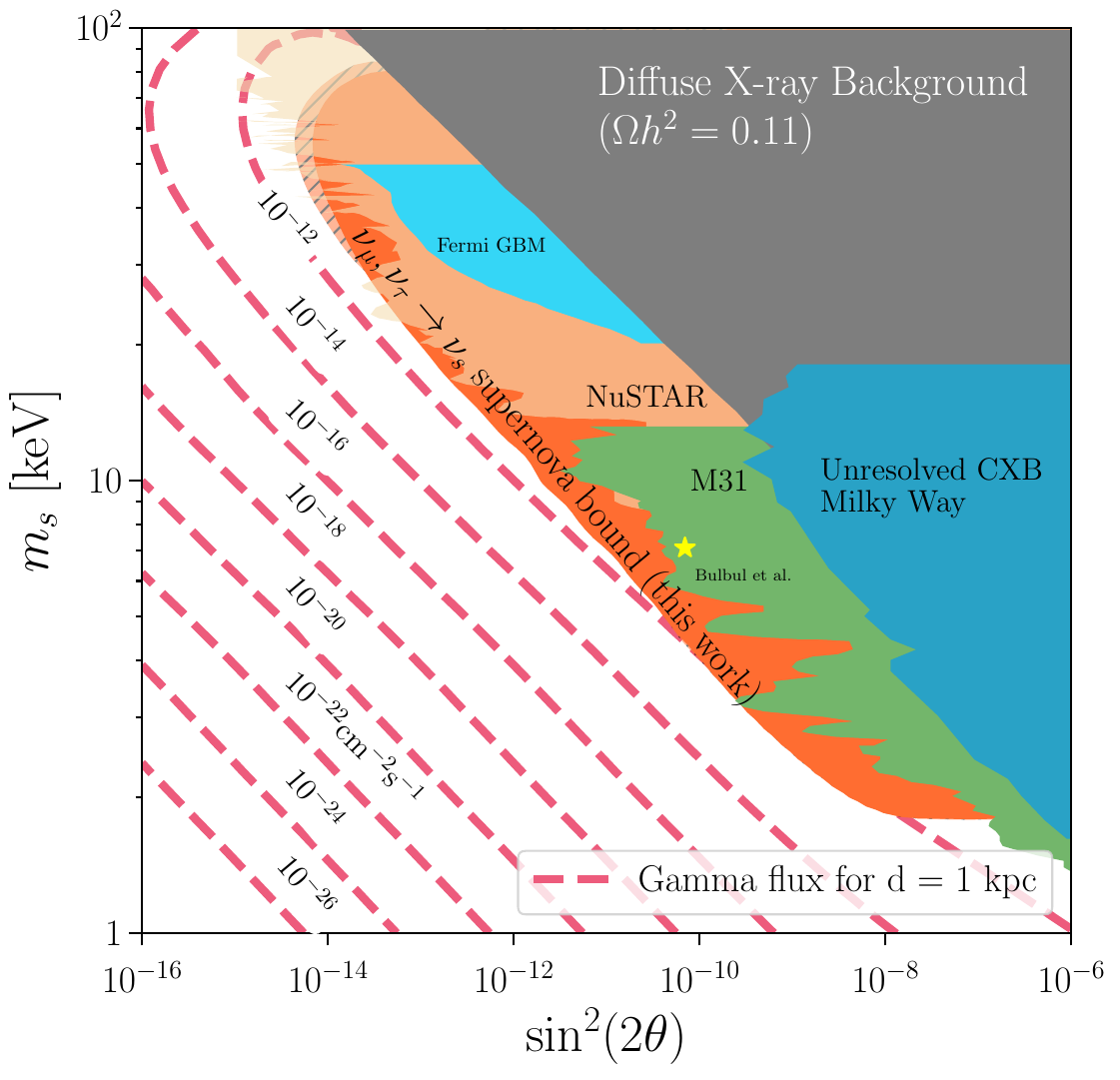}
   \end{tabular}
   \caption{Expected photon flux from the decay of sterile neutrinos produced
     in a supernova at distance $d = 1$~kpc.  We compare to our new energy loss
     limits from \cref{fig:limit-comparison} (orange region) and to constraints
     from X-ray searches~\cite{Horiuchi:2013noa, Abazajian:2006jc,
     Boyarsky:2005us, Abazajian:2009hx, Bulbul:2014sua, Ng:2015gfa}
     (green, dark blue, gray, and light blue regions).
   }
   \label{fig:sensitivity-single-SN}
\end{figure}

The photon flux from $\nu_s$ decay in units of $\text{cm}^{-2}\; \text{sec}^{-1}\;
\text{MeV}^{-1}$ is given by
\begin{align}
  \frac{d\phi_\gamma(E_\gamma)}{dE_\gamma}
    = \frac{2}{4 \pi \, \Delta t \, d^2} \, \bigl[
      1 - \exp(-\tfrac{d}{\gamma} \Gamma_{\nu_s \to \nu_a \gamma})
      \bigr] \frac{d\mathcal{N}_{\nu_s}(E_\nu)}{dE_\nu},
  \label{eq:sn-photon-spectrum}
\end{align} 
where $\gamma = 2E_\gamma / m_s$ is the Lorentz boost factor.  The factor in
brackets accounts for the $\nu_s$ decay probability. Note that the sterile
neutrino spectrum $dN_{\nu_s}(E_\nu) / dE_\nu$ is evaluated at $E_\nu = 2
E_\gamma$.  

We plot $d\phi_\gamma(E_\gamma) / dE_\gamma$ for two sets of benchmark
parameters in \cref{fig:photon-spectrum}, and the total gamma ray flux as a
function of $\sin^2 2\theta$ and $m_s$ is shown in
\cref{fig:sensitivity-single-SN}.
As expected from \cref{fig:spectrum}, and taking into account $E_\gamma = E_\nu
/ 2$, the photon spectrum peaks at $E \sim 10$--$100$~MeV.  Comparing
to observations
\cite{Strong:1994,Kappadath:1996,Kinzer:1999, Hunter:1997,Sreekumar:1997un,
Sreekumar:1997yg, Ackermann:2014usa}, we find that the signal is still several
orders of magnitude below the uncertainty on the background.  A
direct observation of the $\nu_s$-induced photon signal is therefore challenging
even for the next generation of Compton telescopes
\cite{Hunter:2013wla,Moiseev:2015lva,DeAngelis:2017gra} and would require a factor
$\sim 10^5$ improvement compaired to the projected sensitivity of
ComPair~\cite{Moiseev:2015lva} for a supernova at 1~kpc.

\section{Discussion and Conclusions}
\label{sec:conclusion} 

In conclusion, we have computed the flux of hypothetical keV--MeV sterile
neutrinos $\nu_s$ from a supernova.  We have then used our results to constrain
the $\nu_s$ parameter space using energy loss arguments
(\cref{fig:limit-comparison}). Let us summarize once again the assumptions and
approximations on which these limits are based: \emph{(i)} we constrain mixing
of $\nu_s$ with $\nu_\mu$ or $\nu_\tau$, but not with $\nu_e$; \emph{(ii)} we
rely on the supernova simulation from \cite{Huedepohl2010} being a good proxy
for SN~1987A, but in terms of the total energy output and in terms of the
time-dependent matter density and neutrino flux profiles; \emph{(iii)}
we use an approximte treatment of energy transport in the supernova, which is
relevant for the replenishment of neutrinos that have converted into $\nu_s$
(\cref{sec:energy_loss}). While we expect our results to be correct at the
order-of-magnitude level, a more refined treatment, e.g.\ by implementing
$\nu_{\mu,\tau} \to \nu_s$ conversion directly in a hydrodynamical
supernova simulation, is certainly desirabe.  On the other hand, it is
important to note that limits derived from supernovae are independent
of the cosmological abundance of sterile neutrinos, i.e.\ they apply
even to scenarios where $\nu_s$ production in the early Universe is suppressed,
or where $\nu_s$ are diluted after their production.

In the last part of the paper, we have
estimated the gamma ray flux from $\nu_s$ produced in a nearby supernova.
However, we conclude that observing such a signal will be challenging
even for the next generation of Compton telescopes. The signal may come
within reach if sterile neutrinos exist with the parameters suggested by
the 3.5~keV x-ray line found in refs.~\cite{Bulbul:2014sua, Boyarsky:2014jta},
and if a very nearby supernova ($\lesssim 100$~pc) occurs.

Directions for future work include a more detailed calculation of the $\nu_s$
flux from a supernova, going beyond the two flavor approximation and with a
more detailed treatment of collective effects.

\section*{Acknowledgments}

\emph{Acknowledgments.} We are indebted to Georg Raffelt and Ninetta Saviano
for many useful comments on the manuscript, and to Thomas Janka and his
collaborators for providing the data from \cite{Huedepohl2010} in
machine-readable form.  We would also like to thank Markus Ackermann and
John Beacom, Alexei Smirnov, MacKenzie Warren, Shun Zhou for very useful discussions.
VB would like to thank the Wisconsin IceCube Particle
Astrophysics Center (WIPAC) for hospitality.  The work of VB and JK is
supported by the German Research Foundation (DFG) under Grant Nos.\
\mbox{KO~4820/1--1} and FOR~2239 and by the European Research Council (ERC)
under the European Union's Horizon 2020 research and innovation programme
(grant agreement No.\ 637506, ``$\nu$Directions'').  VB is supported by the DFG
Graduate School Symmetry Breaking in Fundamental Interactions (GRK 1581).
Additional support has been provided by the Cluster of Excellence ``Precision
Physics, Fundamental Interactions and Structure of Matter'' (PRISMA -- EXC
1098), grant No.~05H12UME of the German Federal Ministry for Education and
Research (BMBF). CA was supported in part by the National Science 
Foundation (OPP-0236449, NSF-PHY-1505858, and PHY-0969061) and 
by the University of Wisconsin Research Committee with funds granted 
by the Wisconsin Alumni Research Foundation.

\appendix
\section{Depletion of Active Neutrinos}
\label{sec:energy_loss}

In our treatment of adiabatic and collisional production of sterile neutrinos,
we did not yet address in detail the impact of active neutrino depletion on the
$\nu_s$ production rate.  More precisely, when active neutrinos are converted
to $\nu_s$ in the MSW resonance region, their number density in this region
quickly drops.  Simultaneously, neutrinos from elsewhere in the supernova
diffuse into the resonance region, partially compensating the loss due to
$\nu_s$ production.  In the following, we present our method
for taking depletion and diffusion into account.

The differential equations governing active and sterile neutrino energy
transport to and from a small volume element $V$ in a time window $t$ are
\begin{align}
  \frac{1}{V}\frac{dE_a}{dt} &= -\Gamma_{as} + \Gamma_{sa} + I_a - O_a \,,
                                                  \label{eq:E-transport-Ea} \\
  \frac{1}{V}\frac{dE_s}{dt} &=  \Gamma_{as} - \Gamma_{sa} + I_s - O_s \,.
                                                  \label{eq:E-transport-Es}
\end{align}
Here, $E_a$ ($E_s$) is the total energy carried by the active (sterile) neutrinos
in the volume element $V$ at radius $r$, $\Gamma_{as}$
($\Gamma_{sa}$) is the rate of energy transfer by
$\nu_a \to \nu_s$ ($\nu_s \to \nu_a$) conversion, and $I_{s(a)}$ and $O_{s(a)}$
represent the inflow and outflow of energy in sterile (active) neutrinos.
All of these quantities should be understood to be functions of radius $r$ and
time $t$.
We will not include $I_{s(a)}$ and $O_{s(a)}$ directly, but will implement a
simplified treatment of neutrino transport at the very end of the following
derivation. From \cref{eq:boltzmann6}, $\Gamma_{as}$ reads 
\begin{align}
  \Gamma_{as} = \int \! dE \, \frac{\Gamma_x}{2} \ev{P_{\nu_x \to \nu_s}}
                n_x \frac{E^2}{\bar{E}^3} \, f_x(E) E \,.
  \label{eq:Gamma-as-appendix}
\end{align}
$\Gamma_{sa}$ is obtained by replacing the active neutrino energy distribution
and number density with the corresponding quantities for sterile neutrinos.
This term is, however, negligible because the strongest $\nu_s$ production
occurs at the MSW resonance and we can assume that they encounter no more than
one resonance (see discussion below \cref{eq:dNdE-MSW}).  Motivated by the
relation $n_x \bar{E} = E_a / V$, we also define
\begin{align}
  \tilde\Gamma_{as} \equiv \frac{1}{V}
                           \int \! dE \, \frac{\Gamma_x}{2} \ev{P_{\nu_x \to \nu_s}}
                           \frac{E^2}{\bar{E}^3} \, f_x(E) E \,.
  \label{eq:Gamma-tilde}
\end{align}
With this definition, and with the simplification mentioned above,
\cref{eq:E-transport-Ea,eq:E-transport-Es} turn into
\begin{align}
  \frac{1}{V}\frac{dE_a}{dt} &= -\frac{E_a}{\bar{E}} \tilde\Gamma_{as} \,,
                                                  \label{eq:E-transport-Ea-2} \\
  \frac{1}{V}\frac{dE_s}{dt} &=  \frac{E_a}{\bar{E}} \tilde\Gamma_{as} \,.
                                                  \label{eq:E-transport-Es-2}
\end{align}
By solving \cref{eq:E-transport-Ea-2} for $E_a$ and inserting the solution into
\cref{eq:E-transport-Es-2}, we arrive at
\begin{align}
  \frac{1}{V} \frac{dE_s}{dt} = \frac{E_a(r,0)}{\bar{E}} \tilde\Gamma_{as}
    \exp\left[ -\frac{1}{V} \int_{t,\Delta r} \! dV' \, dt' \,
                \frac{\tilde\Gamma_{as}}{\bar{E}} \right] \,,
  \label{eq:E-transport-Es-3}
\end{align}
where $E_a(r,0)$ is energy stored in active neutrinos at radius $r$ at the beginning
of the supernovae cooling phase.

We now model the inflow and outflow terms by convoluting \cref{eq:E-transport-Es-3}
with a Gaussian whose width $\sigma$ represents the scale for active neutrino
transport during the cooling phase:
\begin{align}
  \frac{1}{V} \frac{dE_s}{dt} &= \tilde\Gamma_{as}
    \int_V \! d\hat{V} \frac{E_a(r,0)}{\bar{E}}
    G(r, \mu=\hat r, \sigma=\sigma)  
                                        \nonumber\\
  &\qquad\times
    \exp\left[ -\frac{1}{\hat{V}} \int_{t,\Delta r=\hat{V}} \! dV' \, dt'
                \frac{\tilde\Gamma_{as}}{\bar{E}} \right] \,.
  \label{eq:E-transport-Es-4}
\end{align}
where the gaussian is a truncated gaussian in the $[0,r_{\rm SN}]$ interval given by
\begin{equation}
G(r, \mu, \sigma) = \kappa(\mu,\sigma) e^{\frac{(r-\mu)^2}{2\sigma^2}} \theta(r) \theta(r_{\rm SN}-r).
\end{equation}
with $\kappa(\mu,\sigma)$  the appropriate normalization factor to integrate to unity and $\theta$ the Heaviside function.
We have also implemented an alternative phenomenological procedure in which
the exponential that describes depletion of active neutrinos in
\cref{eq:E-transport-Es-3} is modified. In particular, we include an extra
factor that suppresses the impact of depletion over time:
\begin{align}
  \frac{1}{V} \frac{dE_s}{dt} = \frac{E_a(r,0)}{\bar{E}} \tilde\Gamma_{as}
    \exp\left[ -\frac{1}{V} \int_{t,\Delta r} \! dV' \, dt' \,
      \frac{\tilde\Gamma_{as}}{\bar{E}} \frac{t-t'}{\sigma}\right] \,.
  \label{eq:E-transport-Es-5}
\end{align}
In other words, the depletion factor at time $t$ depends mostly on $\nu_s$
production that happened recently, while the $\nu_a$ that have been converted
to $\nu_s$ longer ago ($t - t' \gtrsim \sigma$) are assumed to have been
replenished by diffusion.

We have verified that we obtain similar results using both techniques. The one
based on \cref{eq:E-transport-Es-5} is more stable and produces fewer numerical
artifacts. The results obtained by applying the phenomenological based procedure, using $\tau$, are presented in
\cref{fig:energy-constraint}.

\section{Hydrodynamical Evolution and Energy Budget}

\begin{figure}
  \centering
  \begin{tabular}{cc}
    \includegraphics[width=0.48\columnwidth]{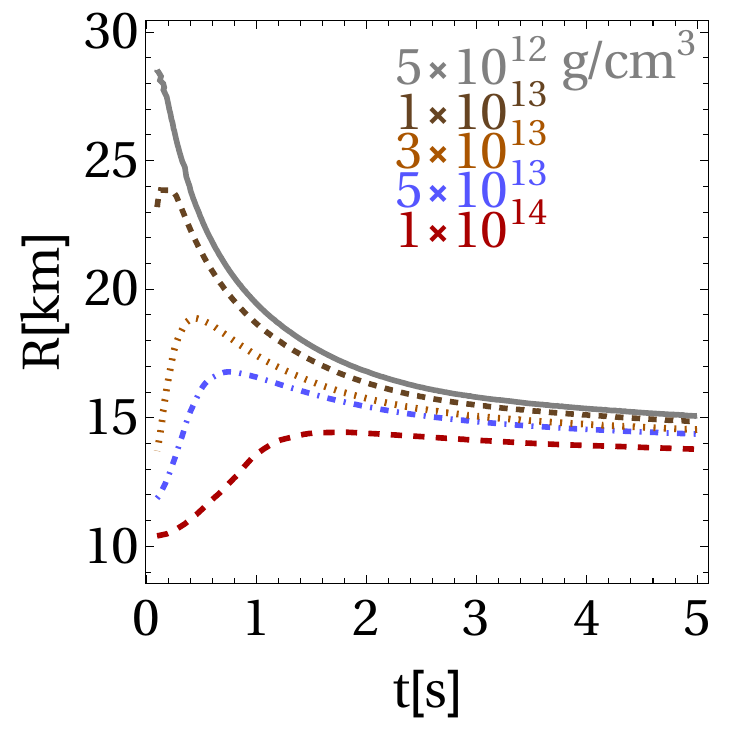} &
    \includegraphics[width=0.48\columnwidth]{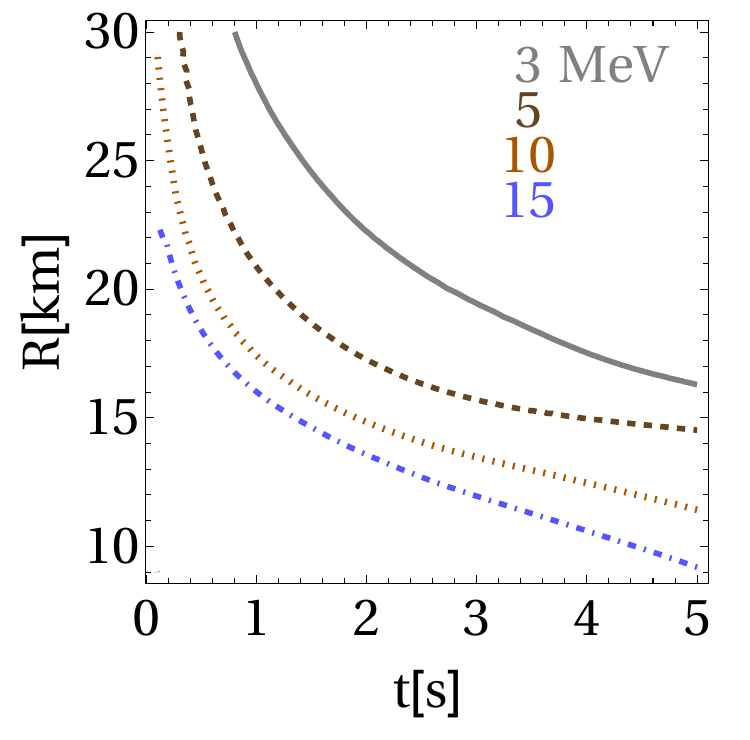} \\
  \end{tabular}
  \caption{\emph{Left:} the supernova density as a function of time $t$ and
    radius $r$.  \emph{Right:} the supernova temperature as a function of $t$
    and $r$. The leftmost edge of the plots at $t=0$ corresponds to the onset
    of the cooling phase.}
  \label{fig:rho-temp}
\end{figure}

In this appendix, we show several ancillary plots illustrating the input and output
of our simulations of sterile neutrino production in supernovae.
In \cref{fig:rho-temp}, we plot the properties of the
supernova simulation from ref.~\cite{Huedepohl2010} on which the results of this
paper are based. In particular, the two panels show contours of constant density
(left) and temperature (right) as a function of the distance from the center of
the star and of the times elapsed since the supernova bounce.
We see that nuclear densities are reached in the inner core, and the supernova
temperature is of order few MeV.

\begin{figure}
  \centering
  \includegraphics[width=0.8\columnwidth]{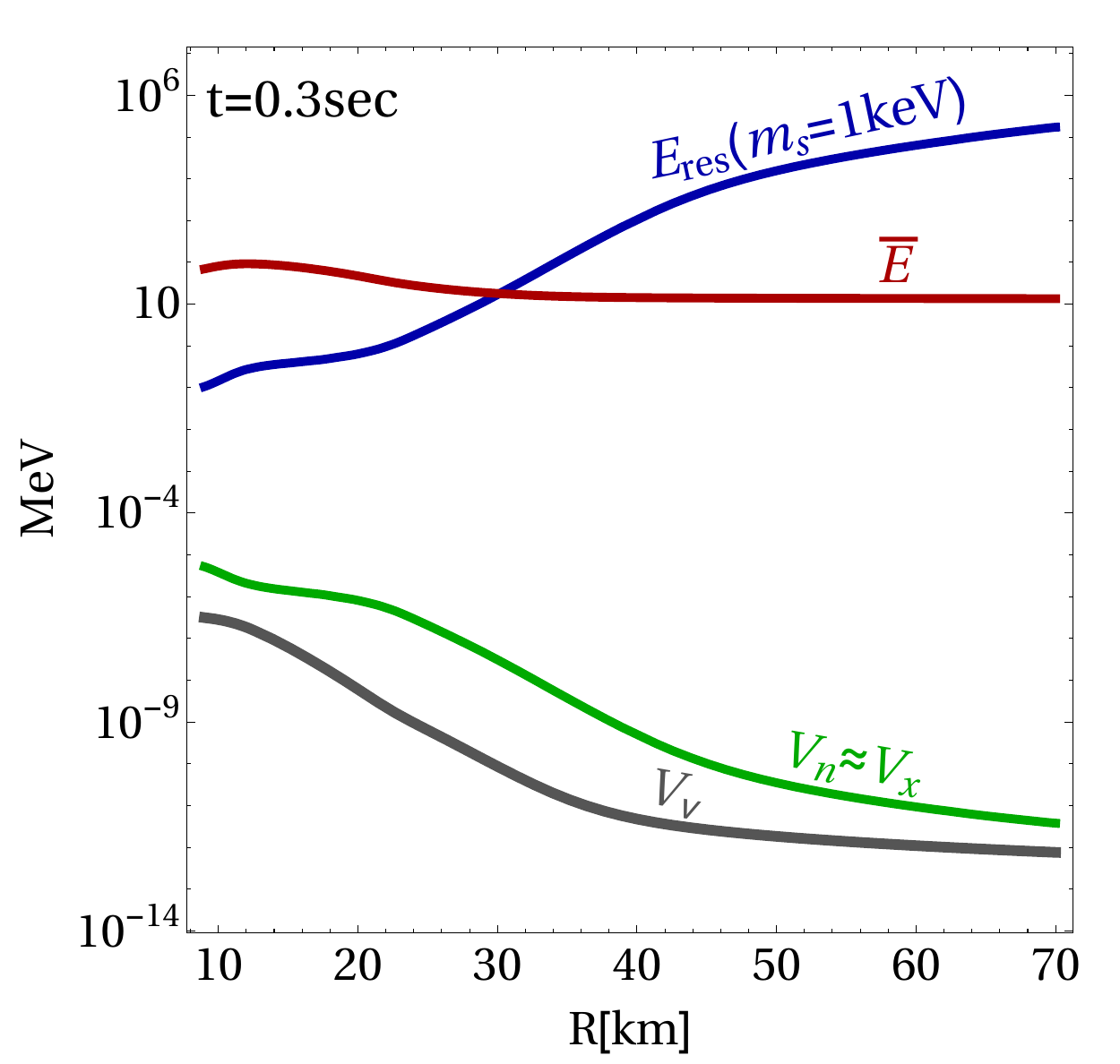}
  \caption{For a specific point in time, $t = 0.3$\,sec, we plot the
    MSW resonance energy $E_\text{res}$ (blue), the average neutrino
    energy $\bar{E}$ (red), and the MSW potential generated by ordinary matter
    $V_n$ (green) and by neutrinos $V_\nu$ (gray) as a
    function of radius $r$.
    }
  \label{fig:v-res}
\end{figure}

Several quantities relevant to neutrino oscillations are plotted in \cref{fig:v-res}
for a specific point in time at $t = 0.3$\,sec after the bounce. In particular,
we compare the energy of the MSW resonance (blue) to the average neutrino energy (red).
The intersection of the two curves indicates that the region favorable for
adiabatic $\nu_x \to \nu_s$ conversion lies at a few tens of kilometers from the
center of the star.  We also show the contributions of coherent forward scattering
on neutrinos $V_\nu$ (gray) and neutrons $V_n$ (green) to the total matter potential
$V_x$.

\begin{figure}
  \centering
  \includegraphics[width=0.8\columnwidth]{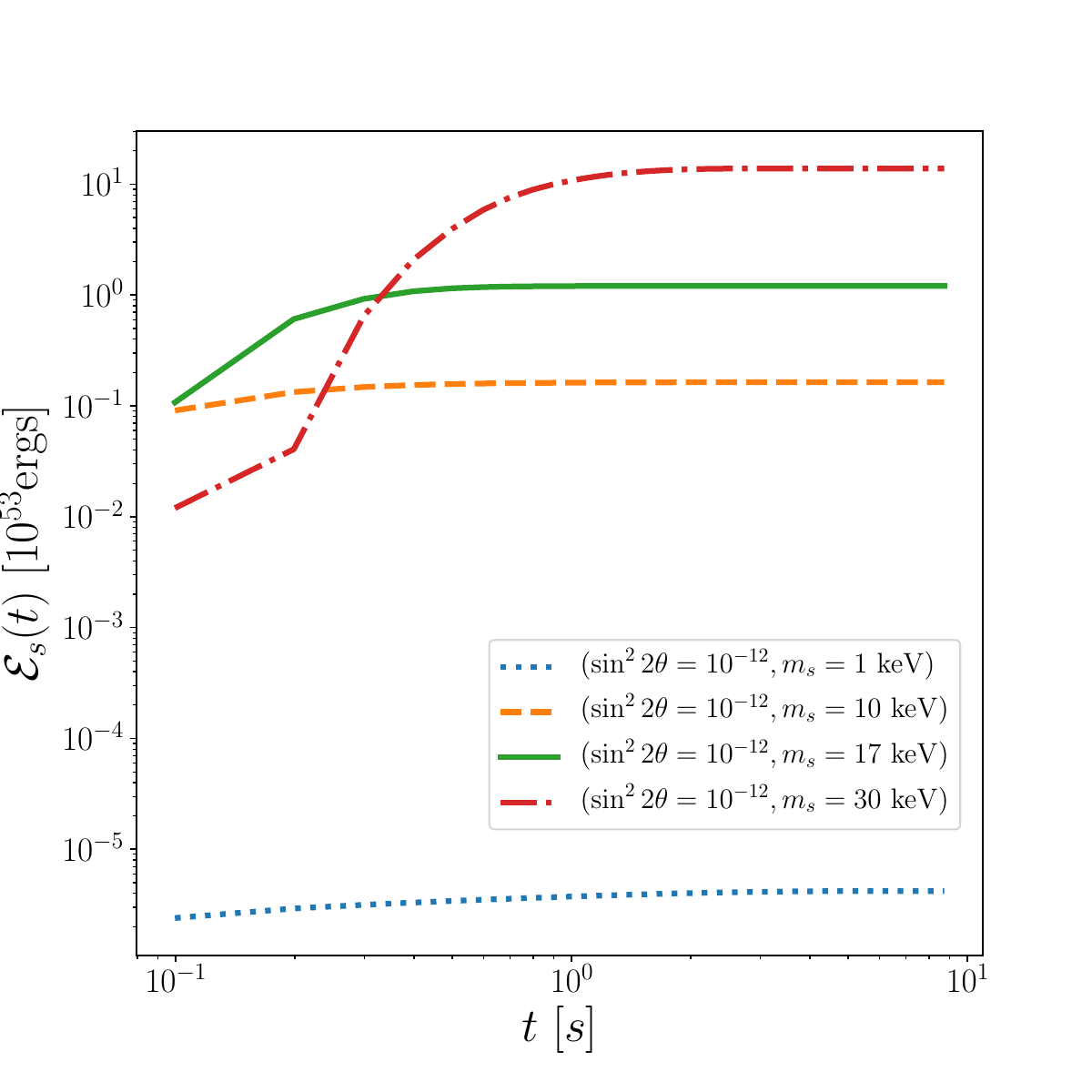}
  \caption{Cumulative energy carried out of the supernovae explosion by
    collisionally produced sterile neutrinos for several parameter points.}
  \label{fig:sterile-energy}
\end{figure}

In \cref{fig:sterile-energy} we show the total integrated energy
$\mathcal{E}_e$ carried out of the star by sterile neutrinos.  We see that such
transport is most efficient during the first second after the bounce.
Comparing the different parameter points shown in \cref{fig:sterile-energy},
we observe that significantly more energy is lost into sterile neutrinos
for $m_s > 10$~keV than for masses below this threshold. The reason is that,
at masses greater than 10~keV, resonant flavor conversion in matter is effective.

\begin{figure}[t]
  \centering
  \includegraphics[width=\columnwidth]{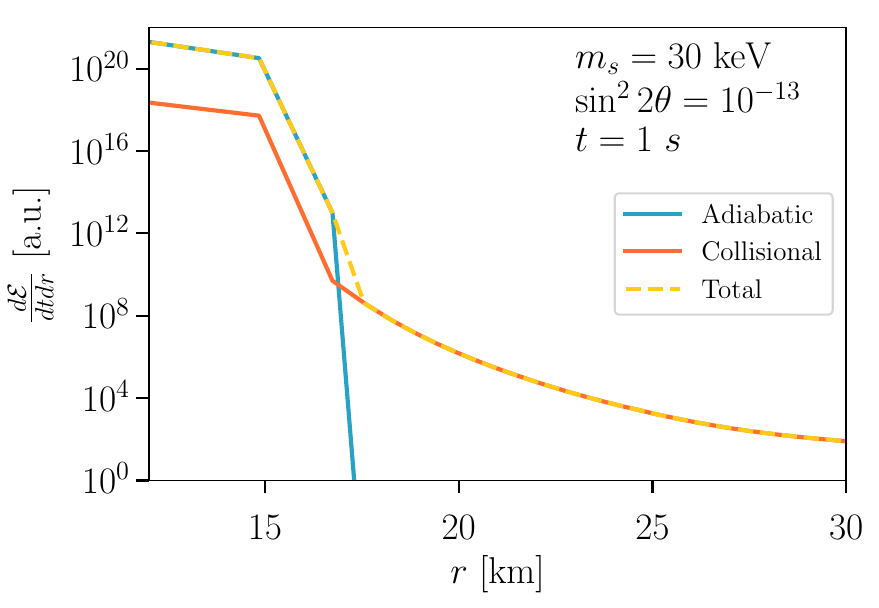}
  \caption{The energy loss rate into $\nu_s$ via both collisional and adiabatic
    conversion for the parameter point $\sin^2 2\theta=10^{-13}$ and $m_s=30$~keV
    at $t=1$ s. Adiabatic contribution dominates at $r\lesssim 15$ km, where the
    resonance condition can be satisfied.}
  \label{fig:energy_cont}
\end{figure}

Finally, in \cref{fig:energy_cont}, we plot the energy loss rate as a function
of radius for a specific parameter point ($\sin^2 2\theta=10^{-13}$,
$m_s=30$~keV). We see that both adiabatic and collisional conversion are
relevant in the inner core ($r \lesssim 15$~km), with the adiabatic contribution
being stronger. At larger radii, adiabatic conversion becomes ineffective because
due to the lower density, resonance energy increases to values outside the
neutrino spectrum.

\bibliographystyle{JHEP}
\bibliography{super-bibliography}

\end{document}